\documentclass[a4paper,twocolumn,11pt,accepted=2021-02-01]{quantumarticle}
\pdfoutput=1
\usepackage[utf8]{inputenc}
\usepackage[english]{babel}
\usepackage[T1]{fontenc}
\usepackage{amsmath}
\usepackage{hyperref}

\usepackage{tikz}
\usepackage{lipsum}

\usepackage[backend=biber,sorting=none,style=phys,biblabel=brackets]{biblatex}
\DeclareUnicodeCharacter{0301}{\'{e}}

\newcommand{\ket}[1]{| #1\rangle}

\newcommand*\colvec[3][]{\begin{pmatrix}\ifx\relax#1\relax\else#1\\\fi#2\\#3\end{pmatrix}}

\newcommand{\qq}{\mathbf{q}}

\newcommand{\um}{$\mu$m}

\renewcommand{\r}{\mathbf{r}}

\begin{document}

\title{Resolution of Quantum Imaging with Undetected Photons}

\author{Jorge Fuenzalida}
\affiliation{Institute for Quantum Optics and Quantum Information, Austrian Academy of Sciences, Boltzmanngasse 3, Vienna A-1090, Austria}
\orcid{0000-0001-7099-1901}
\affiliation{Vienna Center for Quantum Science and Technology (VCQ), Faculty of Physics, Boltzmanngasse 5, University of Vienna, Vienna A-1090, Austria}
\author{Armin Hochrainer}
\affiliation{Institute for Quantum Optics and Quantum Information, Austrian Academy of Sciences, Boltzmanngasse 3, Vienna A-1090, Austria}
\affiliation{Vienna Center for Quantum Science and Technology (VCQ), Faculty of Physics, Boltzmanngasse 5, University of Vienna, Vienna A-1090, Austria}
\author{Gabriela Barreto Lemos}
\affiliation{Instituto de F\'isica, Universidade Federal do Rio de Janeiro, Av. Athos da Silveira Ramos 149, Rio de Janeiro, CP: 68528, Brazil}
\affiliation{Physics Department, University of Massachusetts Boston, 100 Morrissey Boulevard, Boston MA 02125, USA}
\orcid{0000-0003-4901-0309}
\author{Evelyn A. Ortega}
\affiliation{Institute for Quantum Optics and Quantum Information, Austrian Academy of Sciences, Boltzmanngasse 3, Vienna A-1090, Austria}
\affiliation{Vienna Center for Quantum Science and Technology (VCQ), Faculty of Physics, Boltzmanngasse 5, University of Vienna, Vienna A-1090, Austria}
\orcid{0000-0002-4347-7971}
\author{Radek Lapkiewicz}
\affiliation{Institute of Experimental Physics, Faculty of Physics,
University of Warsaw, Pasteura 5, Warsaw 02-093, Poland}
\orcid{0000-0002-7026-7275}
\author{Mayukh Lahiri}
\affiliation{Department of Physics, Oklahoma State University, Stillwater, Oklahoma, USA}
\orcid{0000-0002-7768-424X}
\author{Anton Zeilinger}
\email{anton.zeilinger@univie.ac.at}
\affiliation{Institute for Quantum Optics and Quantum Information, Austrian Academy of Sciences, Boltzmanngasse 3, Vienna A-1090, Austria}
\affiliation{Vienna Center for Quantum Science and Technology (VCQ), Faculty of Physics, Boltzmanngasse 5, University of Vienna, Vienna A-1090, Austria}
\orcid{0000-0002-6778-0887}
\maketitle

\begin{abstract}
Quantum imaging with undetected photons is a recently introduced technique that goes significantly beyond what was previously possible. In this technique, images are formed without detecting the light that interacted with the object that is imaged. Given this unique advantage over the existing imaging schemes, it is now of utmost importance to understand its resolution limits, in particular what governs the maximal achievable spatial resolution.  

We show both theoretically and experimentally that the momentum correlation between the detected and undetected photons governs the spatial resolution — a stronger correlation results in a higher resolution. In our experiment, the momentum correlation plays the dominating role in determining the resolution compared to the effect of diffraction. We find that the resolution is determined by the wavelength of the undetected light rather than the wavelength of the detected light. Our results thus show that it is in principle possible to obtain resolution characterized by a wavelength much shorter than the detected wavelength.
\end{abstract}
\section{Introduction}
Quantum imaging is a rapidly developing field with progress fuelled by developments in photon detection technology \cite{bruschini2019single,lubin2019quantum,nomerotski2019imaging}, nonlinear and quantum optics \cite{Klyshko-1997,Lantz-2005,walborn2010spatial,moreau2019imaging,kutas2020terahertz}, and quantum information theory \cite{Tsang-2016,Demkowicz-2018,magana2019quantum}. Quantum effects applied to conventional imaging techniques have enabled overcoming the classical limits, e.g., shot-noise limit \cite{brida2010experimental,sabines2019twin,garces2020quantum} and diffraction limit \cite{schwartz2013superresolution,classen2017superresolution,unternahrer2018super,tenne2019super}. Besides this, quantum theory has inspired novel imaging techniques such as ghost imaging \cite{strekalov1995observation,pittman1995optical,bennink2002two,gatti2008quantumimaging,aspden2013epr,moreau2019imaging}. Recently, another quantum imaging technique, ``Quantum imaging with undetected photons'' (QIUP), has been introduced \cite{lemos_quantum_2014,lahiri2015theory}. QIUP goes significantly beyond the present quantum imaging capabilities. In this technique, the image is formed without detecting the light that interacted with the object. This imaging technique also inspired a number of applications to infrared spectroscopy \cite{patent,kalashnikov2016infrared}, and optical coherence tomography  \cite{valles2018optical,paterova2018tunable}. 
\par
The concept of QIUP relies on the creation of a photon pair in quantum
superposition at two spatially separated sources \cite{zou1991induced,wang1991induced}. One photon of the pair interacts with the object and is never detected. The image is constructed by detecting the partner photon that does not interact with the object. This distinctive feature of QIUP allows a sample to be probed at a wavelength for which no detector is available, which is an
advantage over existing imaging techniques.
\par
Quantum ghost imaging also uses spatially correlated photon pairs to acquire images \cite{strekalov1995observation,pittman1995optical,bennink2002two,gatti2008quantumimaging,aspden2013epr,moreau2019imaging}. While the achievable resolution of ghost imaging has been studied~\cite{bennink2004quantum,Boyd-twocol-GI-2009,moreau2018resolution}, a detailed analysis for QIUP is hitherto lacking. An understanding of the resolution limits is of utmost importance for assessing  potential applications of QIUP. Since the principle behind QIUP is fundamentally different from principles of all other quantum imaging techniques, a thorough analysis of the resolution is required to understand its resolution limits.
\par
Here we present a detailed analysis of the resolution for QIUP. We address the problem both theoretically and experimentally. We consider the experimental configuration used in Refs. \cite{lemos_quantum_2014,lahiri2015theory}: the object and the camera are located in the far-field of the nonlinear crystals where the photon pairs are emitted. The transverse dimensions of these crystals are much smaller than the diameter of the lenses employed. We find that in this case, the achievable resolution is limited by the momentum correlation between detected and undetected photons. In addition, we show that in this regime, the resolution of QIUP is regulated by the wavelength of the light that probes the object. Our result highlights that the non-degenerate QIUP scheme can be particularly promising for applications where the wavelength of the light probing the object is shorter than the detected wavelength.
\par
In Sec. \ref{sec:theory}, we generalize the theory of imaging presented in Ref. \cite{lahiri2015theory} without the assumption of perfect momentum correlation. In Sec. \ref{sec:exp-mag}, we illustrate the experimental setups and discuss the corresponding magnification. Then in Sec. \ref{Sec:mon-corr-anal}, we present a qualitative discussion of the effect of momentum correlation on resolution. In Sec. \ref{sec:quant-study}, we present the experimental results and compare them with theoretical predictions. In Sec. \ref{Sec:discuss}, we discuss our results, which is followed by proposals of improvements in Sec. \ref{sec:improvement}. Finally, we summarize our results and conclude in Sec. \ref{Sec:conclusions}.

\section{Theory}
\label{sec:theory}
\begin{figure*}[t]
	\centering
	\includegraphics[width=0.8\linewidth]{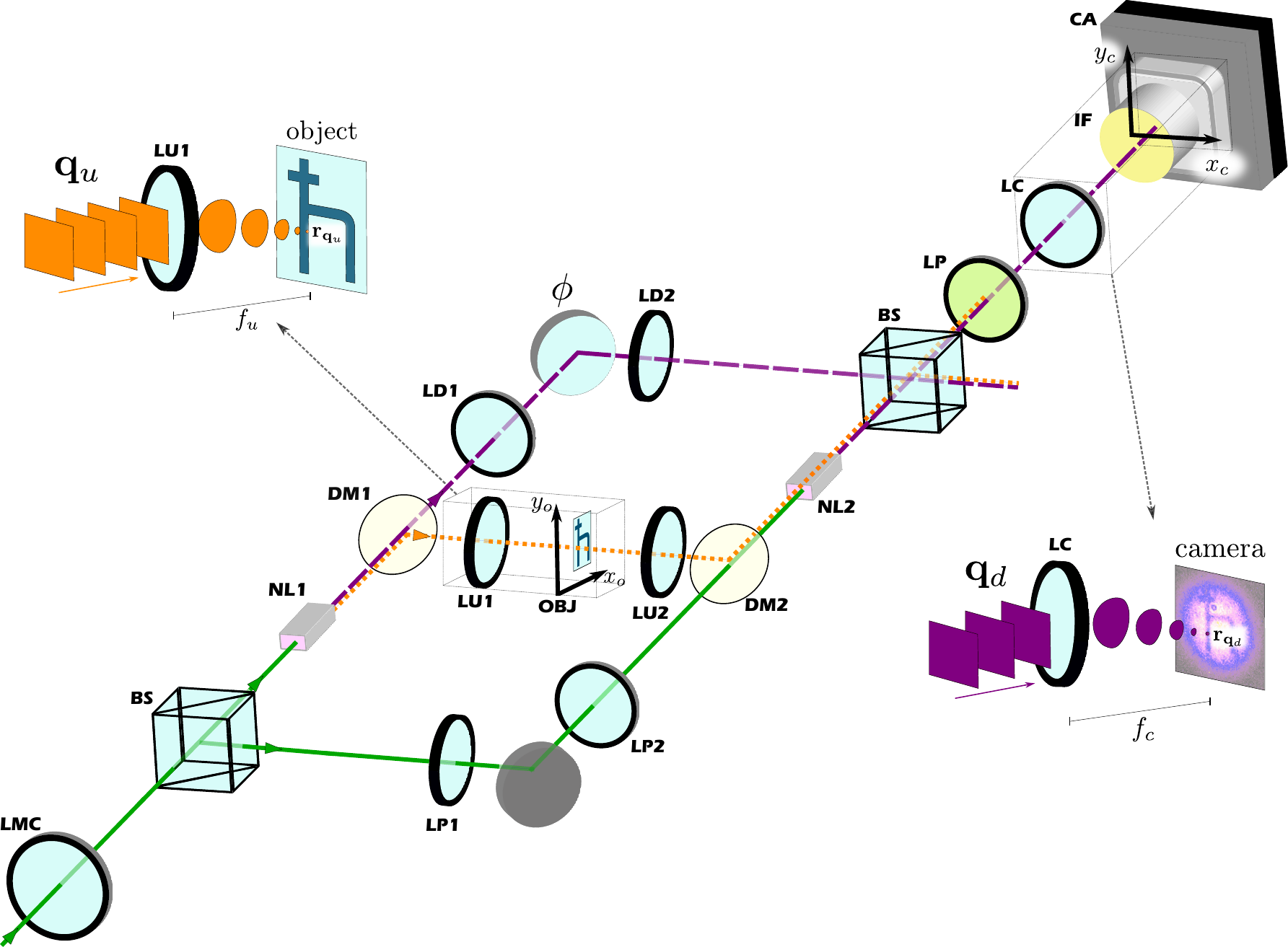}
	\caption{Schematic of Quantum Imaging with Undetected Photons. A photon pair is emitted in one of two identical coherently pumped nonlinear crystals (NL1 and NL2). The pump, the detected beam, and the undetected beam, are represented by the solid green, dashed purple, and spotted orange lines, respectively. The detected beams of both sources are superposed at a beam splitter (BS) and subsequently detected on a camera (CA). The undetected beams are aligned and are never detected. The spatially independent interferometer phase $\phi$ is varied by rotating plate retarders. An object located in the undetected beam between the two sources affects the interference pattern observed on the camera, which allows us to reconstruct both its transmission and phase profiles. Confocal lens systems in the pump beam (LP1 and LP2) ensure equal pump spots at the two crystals, the size of which is determined by the lens LMC. The lenses LU1 and LU2 (focal length $f_u$) in the undetected beam image onto NL2 the spatial modes originating at NL1. Lenses LD1 and LD2 are also arranged for confocal imaging. The camera detects photons after they pass through a long pass filter LP, the lense LC, and an interference filter (IF). Both object and the camera are located in the far-field of the crystals, i.e the transverse momentum of the detected (undetected) beam is mapped to a point on the camera (object) plane, see insets.}
	\label{fig:setup-im-constr}
\end{figure*}

The imaging scheme (Fig. \ref{fig:setup-im-constr}) uses two coherently pumped \emph{identical} nonlinear crystals (NL1 and NL2) as sources of one photon pair ($D$ and $U$) \footnote{In fact it is not necessary to have two identical sources. This imaging scheme can be realized alternatively with a double pass of the laser in a single crystal \cite{patent}}.
The beams of photon $D$ from each source (purple dashed line) are superposed by a beam splitter and subsequently detected on a camera. The beam of the partner photon ($U$; orange dotted line) emitted by NL1 is sent through NL2 and is aligned perfectly with the corresponding beam emitted by NL2
such that the modes of photon $U$ emitted by NL1 and NL2 become identical. As a result of this alignment, which-source information is unavailable, and interference is observed in the camera where only photon $D$ is detected \cite{zou1991induced,wang1991induced}. Photon $U$ is never detected. 
\par
In QIUP, the object is placed in the undetected beam between the two crystals and its complex transmission coefficient appears in the interference pattern \cite{lemos_quantum_2014,lahiri2015theory}.
The absolute value of the transmission coefficient is proportional to the interference visibility \footnote{Visibility, \ensuremath{V=(I_{max}-I_{min})/(I_{max}+I_{min})}, is evaluated at each camera pixel.}.
The phase introduced by the object can be retrieved from the interference pattern.
\par
A detailed theoretical treatment of the imaging scheme was presented in Ref. \cite{lahiri2015theory}, where it was assumed that the transverse momenta of photons $D$ and $U$ are perfectly correlated. 
In another two works, this assumption was dropped and it was suggested that the measured transverse momentum correlation could be associated to the imaging resolution \cite{hochrainer2017quantifying,lahiri2017twin}. Here, we generalize the above-mentioned theoretical and experimental work to rigorously show the key role played by partial momentum correlation in the image resolution.
\par
In our experiment, the photons are emitted into paraxial beams. The photon pair ($D,U$) produced by either of the sources can 
be represented by the normalized quantum state \footnote{Strictly speaking, $\qq_d$ and $\qq_u$ are continuous variables and therefore the sum in Eq. (\ref{source-emits-state}) is to be understood as an integration.}
\begin{align}\label{source-emits-state}
\ket{\psi}=\sum_{\qq_d,\qq_u} C(\qq_d,
\qq_u) \ket{\qq_d}\ket{\qq_u},
\end{align}
where $\qq_d$ and $\qq_u$ represent the transverse wave vectors associated with photons $D$ and $U$, respectively. The transverse momentum correlation of the photon pair is determined by the probability of jointly detecting photon $D$ with transverse momentum $\hbar \qq_d$ and photon $U$ with transverse momentum $\hbar\qq_u$. This joint probability is given by 
$P(\qq_d,\qq_u)=|C(\qq_d,\qq_u)|^2$. 
\par
When the two crystals are weakly pumped, the effect of stimulated emission is negligible \cite{zou1991induced,wang1991induced,wiseman2000induced,lahiri2019nonclassical}. 
The quantum state of a photon pair, produced jointly by the crystals, is given by a linear superposition of the states generated by the crystals individually. The state is, therefore, given by
\begin{align}\label{state-two-source}
\ket{\Psi}=\frac{1}{\sqrt{2}}\sum_{j=1}^{2}\sum_{\qq_{d_j},\qq_{u_j}} C(\qq_{d_j},
\qq_{u_j}) \ket{\qq_{d_j}}\ket{\qq_{u_j}},
\end{align}
where $j=1,2$ labels the two sources (NL1 and NL2) and we have assumed for simplicity that the sources emit with equal probability. Here, we note that a rigorous derivation of Eq. (2) requires considering the vacuum state generated by SPDC (see Ref. \cite{lahiri2015theory}).
\par
Suppose that photon $D$ with transverse momentum $\hbar\qq_d$ is detected at a point, $\r_{\qq_d}\equiv (x_c,y_c)$, on the camera (Fig. \ref{fig:setup-im-constr}). The corresponding partner photon ($U$) with transverse momentum $\hbar\qq_u$ passes through a point, $\r_{\qq_u}\equiv (x_o,y_o)$, on the object. The object is characterized by the complex amplitude transmission coefficient $T(\r_{\qq_u})=|T(\r_{\qq_u})|\exp[i\theta(\r_{\qq_u})]$. 
\par
Following the theoretical steps shown in Ref. \cite{lahiri2015theory} and using the state given by Eq.
(\ref{state-two-source}), one readily finds that the photon counting rate
(intensity) at a point $\r_{\qq_d}\equiv (x_c,y_c)$ on the camera is given by
\begin{align}\label{image-pattern}
I(\r_{\qq_d})= &\sum_{\qq_u} P(\qq_d,\qq_u) \nonumber\\ &\times \left\{1 +|T(\r_{\qq_u})|
\cos[\theta(\r_{\qq_u})+\phi]\right\},
\end{align}
where $\phi$ is a spatially independent interferometric phase that is varied experimentally. An interference pattern is recorded on the camera. It is clear from Eq. \eqref{image-pattern} that both the amplitude and the phase of the transmission coefficient of the object are present in the interference pattern. Furthermore, the presence of
$P(\qq_d,\qq_u)$ in the equation suggests that the resolution depends on the momentum correlation \footnote{Note that Eq. (7) of Ref. \cite{lahiri2017twin} shows the same result for a phase object.}.
\par
We note that the object information is only in the interference term of the expression on the right-hand side of Eq. \eqref{image-pattern}. This term can be used to study the spatial resolution quantitatively. For example, when the object is purely absorptive, we have $\cos(\theta(\r_{\qq_u}))=1$. In this case, it
follows from Eq. \eqref{image-pattern} that the image is represented by the position dependent visibility
\begin{align}\label{visibility-eq} 
V(\r_{\qq_d})= \sum_{\qq_u} p(\qq_u |
\qq_d)|T(\r_{\qq_u})|,
\end{align}
where $p(\qq_u | \qq_d)=P(\qq_d,\qq_u)/\sum_{\qq_u}P(\qq_d,\qq_u)$, for $\sum_{\qq_u}P(\qq_d,\qq_u)\neq 0$, is the conditional probability which depends on the transverse momentum distribution of the pump photon \cite{monken1998transfer}. 
\par
For a phase object (i.e., $|T(\r_{\qq_u})|=1$), the resolution may again be studied from the interference term (see Appendix B).

\begin{figure}[htbp]
	\centering
	\includegraphics[width=0.99\linewidth]{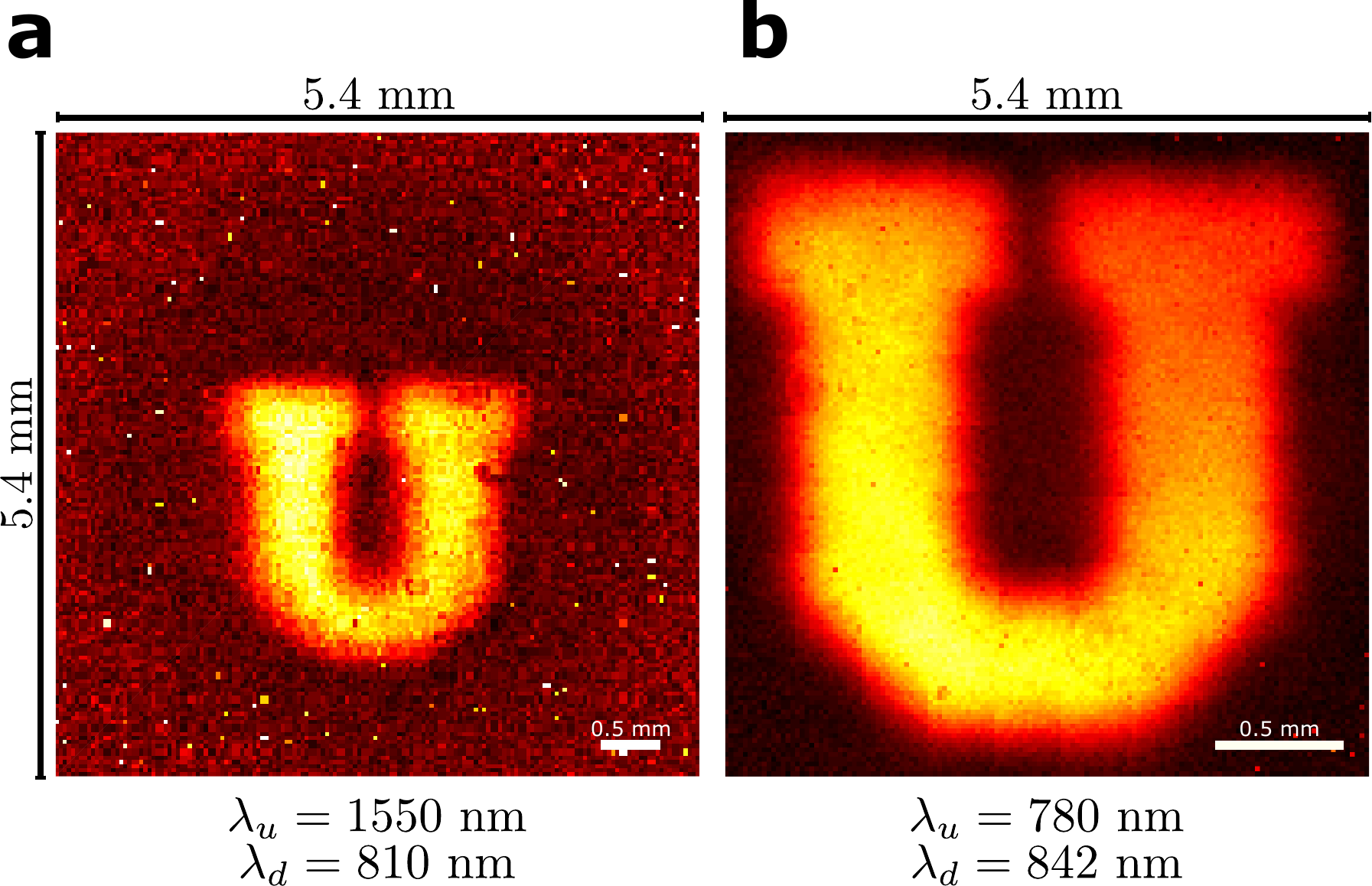}
	\caption{Wavelength dependent image magnification. 
	The same object ($2.417$ mm $\times$ $2.3$ mm) 
	is imaged in two different wavelength configurations, while other parameters such as focal lengths and distances are unchanged between the two scenarios.
    \textbf{a}, Image appearing in $\lambda_d=810$ nm while the object was illuminated with $\lambda_u=1550$ nm (setup 1). Considering the lenses used, the resulting   magnification was $M=1.01\pm0.1$. \textbf{b}, Image appearing in $\lambda_d=842$ nm  while the object was illuminated with $\lambda_u=780$ nm (setup 2). The resulting magnification was  $M=2.16\pm0.1$. $w_p\approx148$ $\mu$m. 
    }
	\label{fig:magnification}
\end{figure}

\section{Experimental Setups and Magnification}\label{sec:exp-mag}

We build two experimental setups characterized by two sets of wavelengths. For setup 1, $\lambda_d=810$ nm and $\lambda_u=1550$ nm, i.e., the undetected photon's wavelength is longer than that of the detected photon. For setup 2, $\lambda_d=842$ nm, $\lambda_u=780$ nm, i.e., the undetected photon has the shorter wavelength. All other parameters of the imaging system (distances, focal lengths, etc.) are the same in the two setups. This also includes the distance between object and camera. The focal length of positive lenses, LU1 and LU2 (Fig. \ref{fig:setup-im-constr}), placed in the path of the undetected photon is $f_u=75$ mm. The positive lens, LC, placed in front of the camera (Fig. \ref{fig:setup-im-constr}) has focal length $f_c=150$ mm.
\par
The image is constructed by varying the spatially independent interferometer phase, $\phi$. Subsequently, phase shifts and the visibility of the interference observed on the camera are evaluated using sinusoidal fits to the intensity data for each pixel individually. The experimental study of the resolution considers purely absorptive objects. The same conclusion is expected for a phase object, as we demonstrate in Appendix B. 
\par
An important feature of the imaging scheme is that the magnification ($M$) depends on the wavelengths of both photons. It has been shown that the magnification is given by \cite{lahiri2015theory,lemos_quantum_2014}
\begin{align}\label{mag-form}
M=\frac{f_c\lambda_d}{f_u \lambda_u}.
\end{align}
We experimentally verify this formula by imaging the same object using the two setups (Fig. \ref{fig:magnification}). We find that $M_1= 1.01\pm0.1$ (Fig. \ref{fig:magnification}a) and $M_2=2.16\pm 0.1$ (Fig. \ref{fig:magnification}b) for setup 1 and setup 2, respectively. The experimentally measured values agree with the theoretical predictions. Using these magnification values, we estimated the far field illumination area. We obtained $10.4$~mm$^2$ for setup~1 and $6.9$~mm$^2$ for setup~2.
\par
The wavelength dependence of the magnification plays an important role in understanding the resolution of the imaging system as we point out later.

\section{Resolution and Momentum Correlation}\label{Sec:mon-corr-anal}
The analysis presented in Sec. \ref{sec:theory} (in particular Eqs. \eqref{image-pattern} and \eqref{visibility-eq}) suggest that a stronger momentum correlation between the two photons leads to a better resolution. We now provide a physical explanation and test the theoretical understanding by analyzing images of the same object taken with different strengths of momentum correlation.
\begin{figure}[htbp]
	\centering
	\includegraphics[width=1\linewidth]{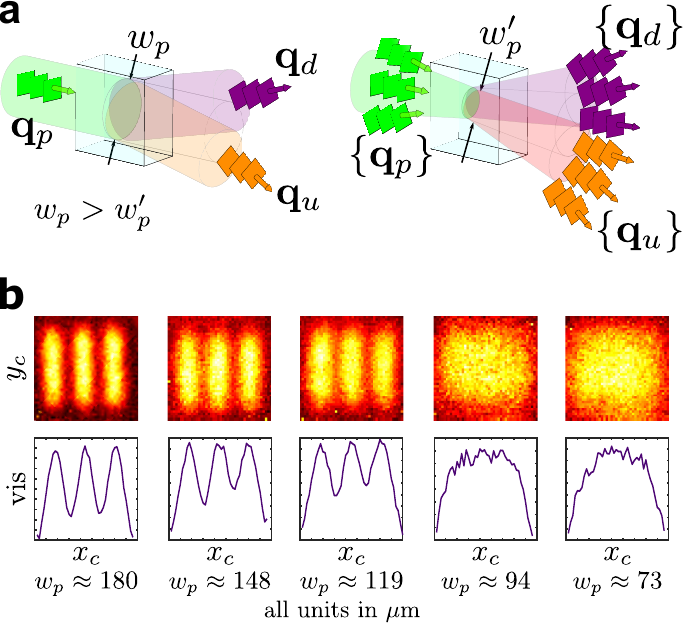}
	\caption{
    Dependence of the resolution on the transverse momentum correlation of photon pairs produced at NL1 and NL2.	
    In (a), photon pairs produced by SPDC exhibit a stronger momentum correlation if the crystal is illuminated by a large pump waist $w_p$ compared to a narrow pump focus $w_p'$, composed of many plane wave components. In (b), images of a resolution target (slit width= 250 \um) obtained by the method of Quantum Imaging with Undetected Photons are depicted for different values of $w_p$. It can be seen that a larger pump waist (stronger momentum correlation) is associated with a better imaging resolution.}
	\label{fig:mom-corr-pictures}
\end{figure}
\par
Both the object and the camera are located in the far-field (Fourier plane) of the nonlinear crystals (Fig. \ref{fig:setup-im-constr}). If a photon with transverse momentum $\hbar\qq_d$ is detected at a point $\r_{\qq_d} \equiv (x_c,y_c)$ on the camera, it follows from the lens formula that $\r_{\qq_d} \approx f_c\lambda_d \qq_d /(2\pi)$ \cite{lahiri2015theory}. In the ideal case of perfect momentum correlation, i.e., when $p(\qq_u|\qq_d)\propto \delta(\qq_d+\qq_u)$, the detection of a photon ($D$) with momentum $\hbar \qq_d$ allows us to infer with certainty that the momentum of its undetected partner ($U$) is $\hbar \qq_u$. Since the object is also located at the Fourier plane of the crystals, the undetected photon ($U$) with transverse momentum $\hbar \qq_u$ arrives at a unique point, $\r_{\qq_u} \approx f_u\lambda_u \qq_u /(2\pi)$, on the object plane. Therefore, the interference pattern recorded at point $\r_{\qq_d}$ will contain information about only one point $\r_{\qq_u}$ on the object when the momentum correlation is perfect. For example, when the transmission coefficient at $\r_{\qq_u}$ is zero, the visibility measured at $\r_{\qq_d}$ also becomes zero. However, if the momentum correlation is not perfect, one point on the camera does not correspond to only one point on the object. This is because the momentum ($\hbar \qq_u$) of photon $U$ now varies over a range
when the photon $D$ is detected with a particular momentum. This range becomes wider with a weaker momentum correlation. Therefore, the information from a range of points on the object appears at a single point on the camera, i.e., a weaker momentum correlation reduces the spatial resolution.
\par
In the experiment, we use Gaussian pump beams and focus them on the two crystals in such a way that the waists of the two pump beams are equal. The conditional probability governing the momentum correlation is given by \cite{monken1998transfer}
\begin{align}\label{cond-prob-form}
p(\qq_u|\qq_d)\propto &\exp\left(-|\qq_d+\qq_u|^2 w_p^2/2\right)\nonumber \\ &\approx \exp\left(-|\qq_p|^2 w_p^2/2\right),
\end{align}
where $w_p$ is the waist of a pump beam, and $\qq_p$ is the transverse component of the pump wave vector. It follows that a stronger correlation is obtained for a larger beam waist (Fig. \ref{fig:mom-corr-pictures}a). We change the correlation by varying the waist of the two pump beams simultaneously by lens LMC \cite{hochrainer2017quantifying,lahiri2017twin}.
\par
Figure \ref{fig:mom-corr-pictures}b presents experimentally obtained images (visibility at each point on the camera) of a resolution target obtained with different pump waist sizes for $\lambda_u=1550$ nm and $\lambda_d=810$ nm. 
The line of the object has a width of 250 $\mu$m (see Appendix \ref{app-testtarge} for object details). The results clearly show that resolution reduces with a weaker momentum correlation.

\section{Quantitative Study and Dependence on Wavelength}\label{sec:quant-study}

\begin{figure*}[ht]
	\centering
	\includegraphics[width=0.9\linewidth]{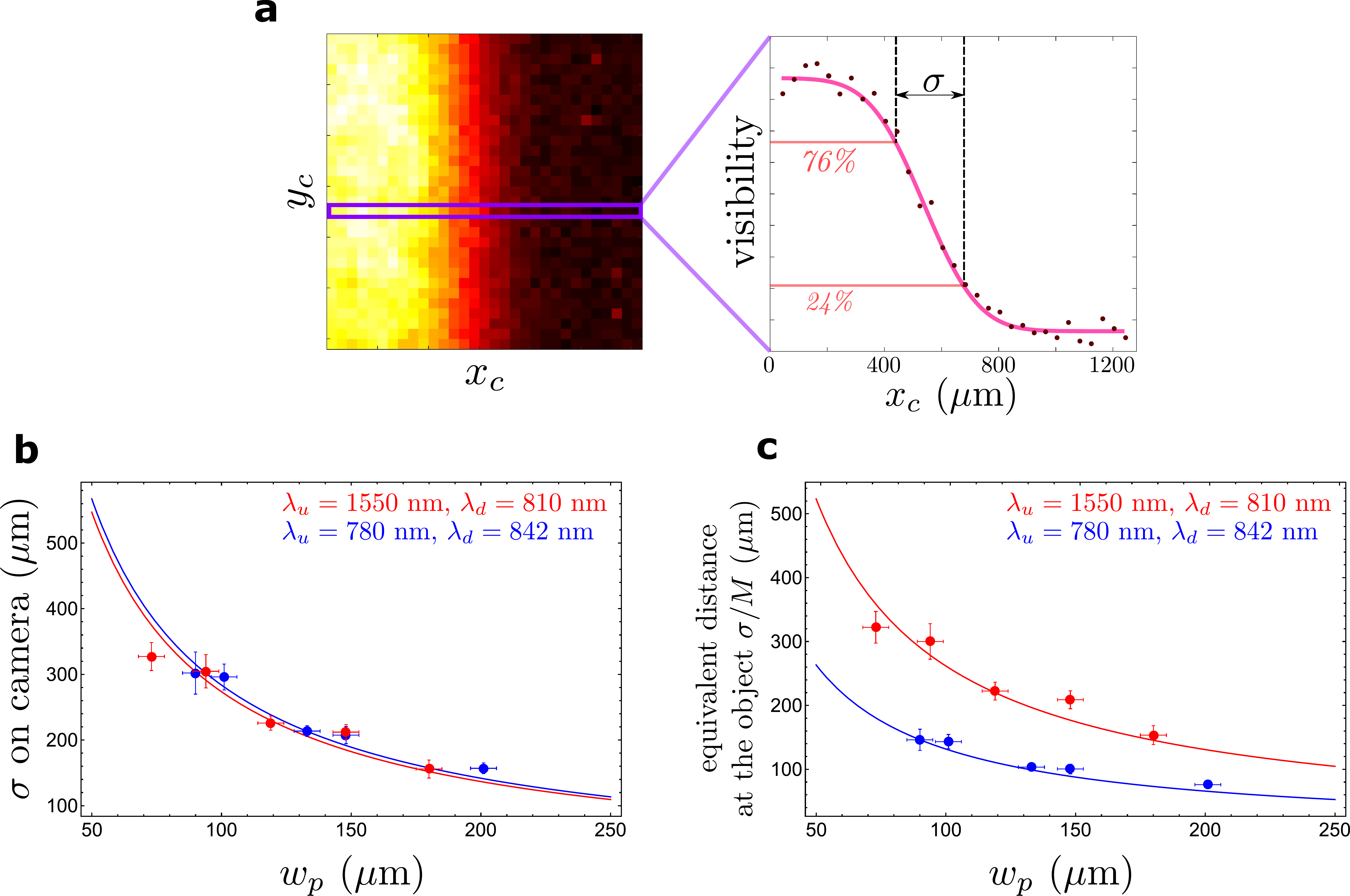}
	\caption{Quantitative measurement of the resolution $\sigma$ in two pairs of wavelengths and varying the pump waist (momentum correlation). We evaluate visibility images of a sharp edge and determine a characteristic distance $\sigma$ of the transition (24\% to 76\%) between minimum and maximum visibility. This is done by fitting an error function to the visibility cross section of the image (a), see text. 
 The plots (b) and (c) show the results at two sets of wavelengths, $\lambda_u=1550$ nm, $\lambda_d=810$ nm (red), and  $\lambda_u=780$ nm, $\lambda_d=842$ nm (blue). In (b), the directly evaluated distance $\sigma$ as it appears on the camera is plotted and compared to the theoretical predictions (solid lines). Under reasonable assumptions it is equivalent to the width of a point spread function, i.e. to the spot size on the camera produced by a point like object. In (c), this distance is scaled with the respective magnifications ($\sigma/M$), which provides a quantitative measure for the resolvable feature size on the object.
  In both cases, the resolution is reduced for a smaller pump waist, i.e., a stronger momentum correlation leads to a better resolution. The data further show that while the absolute size of the blur on the camera is given by the detected wavelength, the associated resolution is given by the wavelength of the undetected light.}
	\label{fig:esf-resultati}
\end{figure*}
We now study the connection between resolution and momentum correlation quantitatively. We also address the question of which wavelength determines the resolution.
\par
We first consider the edge-spread function (ESF) that allows us to study the unsharpness of the imaging system quantitatively \cite{harms1977new}. A knife-edge is
used as an object. The ESF is obtained from the image of the knife-edge. Because we use spherical lenses throughout the setup, we assume axial symmetry. We place the knife-edge parallel to the
$y_o$ axis and along the line $x_o=x_o^{\prime}$, so that 
the undetected beam is blocked between the crystals in the region $x_o<x_o'$ \footnote{The knife-edge is theoretically modeled by the Heaviside step function, which has the form $T(\r_{\qq_u})\equiv T(x_o,y_o)=0$ for $x_o<x_o'$ and $T(\r_{\qq_u})\equiv T(x_o,y_o)=1$ for $x_o \geq x_o'$.}. It now follows from Eqs. (\ref{visibility-eq}) and (\ref{cond-prob-form}) that \footnote{We replace the sum by an integration and use the relations $\r_{\qq_u}=f_u\lambda_u \qq_u /(2\pi)$ and $\r_{\qq_d} \approx f_d\lambda_d \qq_d /(2\pi)$.} the ESF is given by $V(x_c)\propto 1-\text{erf}[(x_c-Mx'_o)/\sigma]$, where erf is
the error function, $M$ is the magnification, and $\sigma$ is given by
\begin{align}\label{PSF-width}
\sigma=\frac{f_c\lambda_d}{\sqrt{2}\pi w_p}.
\end{align} 
It can be checked from the formula of ESF that $\sigma$ is the distance on the camera for which the value of ESF (visibility) rises from approximately $24\%$ to $76\%$ of the maximum attainable value. Experimentally, the quantity $\sigma$ is determined from the image (position dependent visibility) of the knife-edge (Fig. \ref{fig:esf-resultati}a). 
\par
In Fig. \ref{fig:esf-resultati}b, we plot the experimentally measured values of $\sigma$ (data points with error bars) against the pump waist ($w_p$) for both setup 1 and setup 2. The theoretically predicted results (solid lines) are also given for comparison. All measured data differ by less than two standard deviations from the corresponding theoretical values \footnote{The experimental error (standard deviation) was obtained from the sinusoidal fit of the experimental data, and was propagated through the visibility calculations.}. It is clear that $\sigma$ increases as the pump waist ($w_p$) decreases. We discussed in Sec. \ref{Sec:mon-corr-anal} that the momentum correlation decreases as the pump waist decreases. Therefore, Fig. \ref{fig:esf-resultati}b demonstrates that the resolution decreases as the momentum correlation reduces.
\par
We note from Eq. \eqref{PSF-width} that the quantity $\sigma$ does \emph{not} depend on the wavelength ($\lambda_u$) of the light that illuminates the object. It only depends on the wavelength ($\lambda_d$) of the detected light. The experimental results (Fig. \ref{fig:esf-resultati}b) confirm this fact: although the values of $\lambda_u$ are widely different (1550 nm and 780 nm) for the two setups, the values of $\sigma$ are relatively close because the values of $\lambda_d$ are also close (810 nm and 842 nm) for the two setups. However, this does not mean that the resolution is characterized by the wavelength of the detected light. The reason is that the quantity $\sigma$ represents a distance measured on the camera, whereas resolution refers to minimum resolvable distance at the object. In order to obtain a measure of spatial resolution, one needs to divide $\sigma$ by the magnification \cite{boreman2001modulation}. An appropriate measure of resolution for our experiment is therefore
\begin{align} \label{PSFscaled}
\frac{\sigma}{M}=\frac{f_u\lambda_u}{\sqrt{2}\pi w_p},
\end{align}
where $M$ and $\sigma$ are given by Eqs. (\ref{mag-form}) and (\ref{PSF-width}), respectively. 
We find that $\sigma/M$ depends only on the wavelength of the undetected photon, i.e., the light illuminating the object. The experimentally obtained values of $\sigma/M$ are shown and compared with the theoretical prediction in Fig. \ref{fig:esf-resultati}c. The results show that for the same pump waist at the crystal and consequently for the same strength of the momentum correlation (see Eq. \eqref{cond-prob-form}), the resolution is significantly better when the wavelength of the light illuminating the sample is shorter. This holds despite the fact that the wavelength (842 nm) of the detected light is longer than that in the other case (810 nm). Figure \ref{fig:esf-resultati}c strongly suggests that the wavelength of the undetected photon characterizes the resolution.

\begin{figure}[t]
	\centering
	\includegraphics[width=1\linewidth]{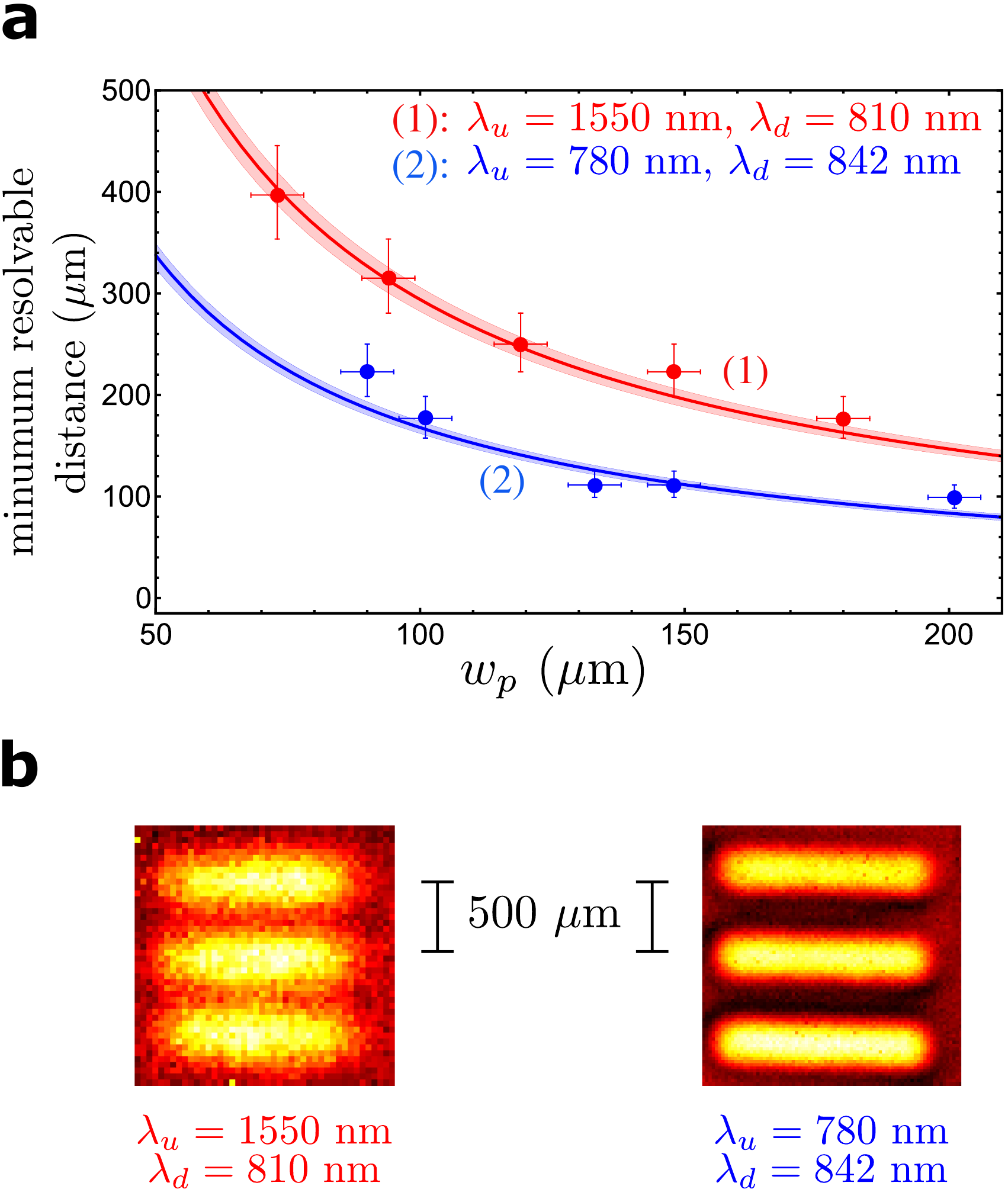}
	\caption{Measurement of resolution using the 1951 USAF resolution test chart. Minimum resolvable distance is defined as the ``width of 1 line in $\mu$m'' when two consecutive slits are just resolved (see main text). \textbf{a}, minimum resolvable distance is plotted against pump waist ($w_p$) for the two experimental setups. The experimental data points match closely with the theoretical predictions (solid lines with shaded area that simulates possible error). When the object is illuminated by light of lower wavelength (curve (2)), the resolution is better even if that light is not detected. \textbf{b}, Image of a particular collection of slits for two experimental setups with an identical $w_p\approx148$  $\mu$m. The width of 1 line in micrometers is 250 $\mu$m. Clearly, when the object is illuminated with the shorter wavelength, the resolution is much higher.}
	\label{fig:target-results}
\end{figure}
\par
We now perform a direct test of the wavelength dependence of the resolution using the 1951 USAF resolution test chart (Fig. \ref{fig:target-results}). We experimentally obtain images (position dependent visibilities) of different parts of the test target. Images (position dependent visibility) of two consecutive slits of the test chart overlap and form a double-humped distribution (see Appendix C). In analogy to the Rayleigh criterion (\cite{born2013principles}, Sec. 7.6.3), we assume that two consecutive slits are just resolved when the ratio ($R$) of the visibility at the dip (minimum) to that at a peak (maximum) is equal to $0.81$. Consequently, we choose the minimum resolvable distance as the ``width of $1$ line in $\mu$m'' when two consecutive slits are just resolved.
\par
In Fig. \ref{fig:target-results}a, we plot the experimentally measured minimum resolvable distances against the pump waist (for both setups) and compare them with the theoretical predictions. Data points (filled circles with error bars) and the numerically simulated curves labeled by (1) and (2) correspond to setup 1 ($\lambda_u=1550$ nm, $\lambda_d=810$ nm) and setup 2 ($\lambda_u=780$ nm, $\lambda_d=842$ nm), respectively. Since the width of $1$ line in $\mu$m cannot be varied continuously in the 1951 USAF resolution test chart, the experimentally measured values of $R$ are never exactly $0.81$. We choose the slits for which $R$ is less than and nearest to 0.81. The theoretical predictions (solid lines with shaded areas) are made considering the mean values and standard deviations of the experimentally measured $R$ ($0.70\pm 0.04$ for setup 1 and $0.74\pm 0.03$ for setup 2). In Fig. \ref{fig:target-results}b, we show images of the same set of slits obtained by the two imaging setups for a fixed pump waist ($w_p\approx148$  $\mu$m). Our results clearly demonstrate that the image resolution is higher if the object is probed using a shorter wavelength, although the detected wavelength is longer in this case.

\section{Discussion}\label{Sec:discuss}
\begin{figure}[t]
	\centering
	\includegraphics[width=1 \linewidth]{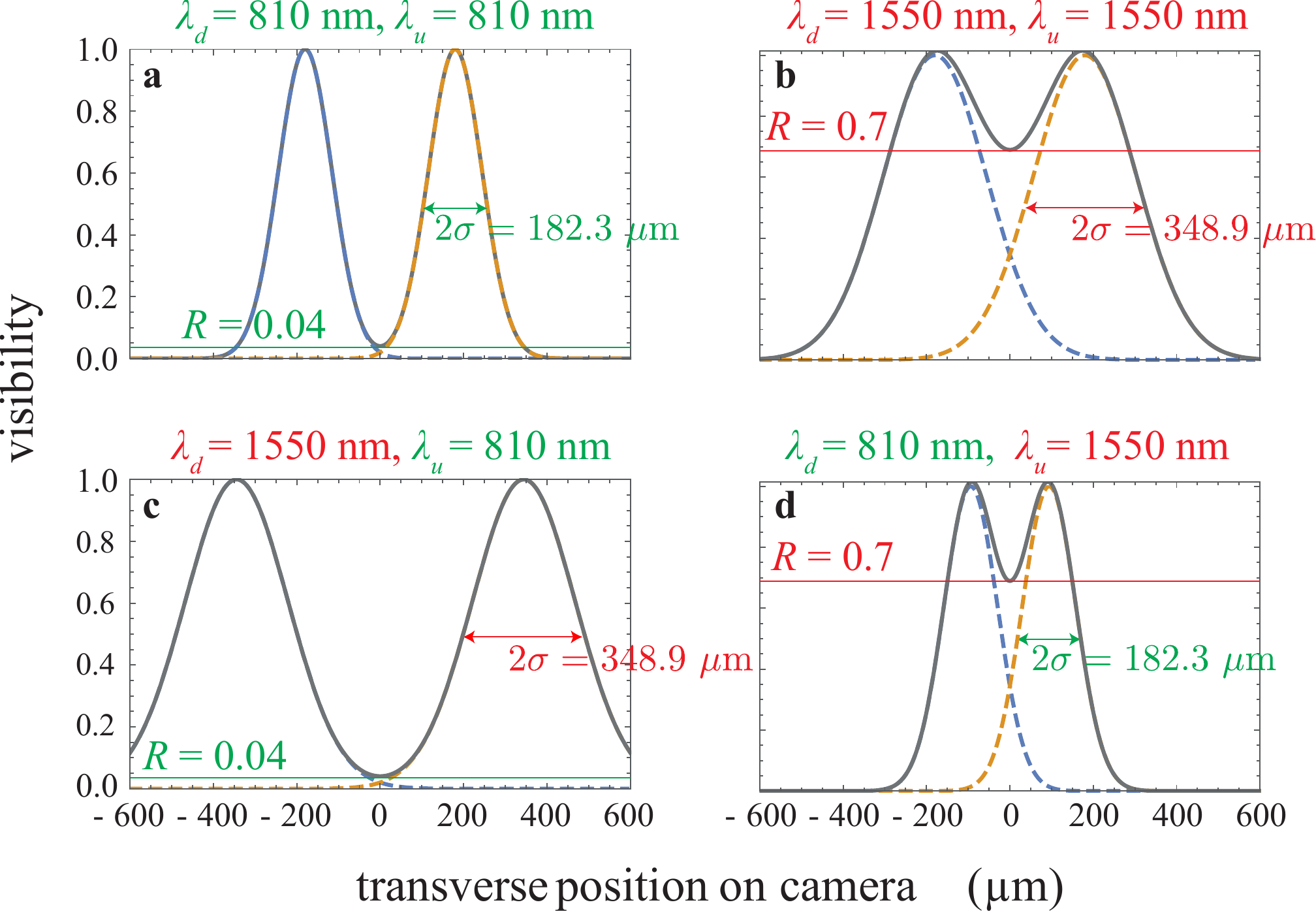}
	\caption{
	Theoretical calculations to illustrate the  role of the two wavelengths in resolution. The resolution is given by the wavelength of the light that illuminates the object ($\lambda_u$), while the absolute size of the image of a point object is determined by the detected wavelength ($\lambda_d$). The transverse visibility image on the camera is simulated for two points separated by $d=180$ $\mu$m. \textbf{a}, Both wavelengths are given by $\lambda_d=\lambda_u=810$ nm. \textbf{b}, Both wavelengths are $1550$ nm. The two points produce images on the camera with widths given by the longer wavelength. As expected, the two slits are less resolvable, as the contrast between them decreases ($R=0.7$). \textbf{c}, The object is probed with $\lambda_u=810$ nm and the image is obtained in the detected beam with $\lambda_d=1550$ nm. Although the width of a spot produced by an individual slit is given by the longer wavelength, the contrast between the two slits is the same as in A ($R=0.04$). That is, the resolution is governed by $\lambda_u=810$ nm. This is possible because the wavelength ratio leads a magnified image on the camera. \textbf{d}, The object is probed with $\lambda_u=1550$ nm and the image is obtained by detecting light at $\lambda_d=810$ nm. This figure shows that the width of the individual spots is given by the detected wavelength, whereas the resolution is governed by the undetected wavelength.} 
	\label{fig:signalandidlerwavelengthdependence}
\end{figure}
Our results show that the width ($\sigma$) of the edge-spread function on the camera is determined by the wavelength of the detected light, while the resolution is determined by the wavelength of the undetected light that illuminates the object. The roles of the two wavelengths can be summarized by analyzing the simulated image of a pair of points in different wavelength configurations. Figure \ref{fig:signalandidlerwavelengthdependence} depicts the computed visibility cross section (image) on the camera for an object that consists of two points with a separation $d=180$ $\mu$m. Focal lengths and distances are assumed to be the same as in our experiment and the pump waist is $w_p=300$ $\mu$m. 
\par
The image of an individual point turns out to be a Gaussian distribution of visibility. This distribution effectively represents the point-spread function for our imaging system. Since we have assumed axial symmetry, the width of this distribution on the camera ($2\sigma$)
is directly related to the width of the edge-spread function ($\sigma$). For a pair of points, a double humped distribution is formed (Fig. \ref{fig:signalandidlerwavelengthdependence}). The ratio of the visibility at the minimum to the visibility at a peak is denoted by $R$. The lower the value of $R$, the better we can resolve the two points. 
\par
In Fig. \ref{fig:signalandidlerwavelengthdependence}a, the wavelengths of detected and undetected photons are $\lambda_d=\lambda_u=810$ nm, whereas in Fig. \ref{fig:signalandidlerwavelengthdependence}b both wavelengths are $\lambda_d=\lambda_u=1550$ nm. Since the magnification is linearly proportional to the ratio $\lambda_d/\lambda_u$ (see Eq. \eqref{mag-form}), it has the same value for these two cases. The width of the point-spread function ($2\sigma$) on the image is given by the respective detected wavelength, which results in a better contrast between the two points in the case of the shorter wavelength (Fig. \ref{fig:signalandidlerwavelengthdependence}a; $R=0.04$) compared to the longer wavelength (\ref{fig:signalandidlerwavelengthdependence}b; $R=0.7$). Figure \ref{fig:signalandidlerwavelengthdependence}c shows a situation in which the same object is imaged with $\lambda_d=810$ nm and $\lambda_u=1550$ nm, i.e., the shorter wavelength is used to probe the object. While the width of the point spread function on the image is given by the detected wavelength, the wavelength dependent magnification results in a contrast between the two points of the same value ($R=0.04$) as in Fig. \ref{fig:signalandidlerwavelengthdependence}a. Thus, the resolution can in principle exceed the limit of the detected wavelength and is instead given by the undetected wavelength. The opposite case is depicted in Fig. \ref{fig:signalandidlerwavelengthdependence}d: although the width of the point spread function on the camera is determined by the shorter detected wavelength, the contrast between the two points attains the same value ($R=0.7$) as if only the longer wavelength were used. This is due to the fact that the wavelength ratio leads to a demagnification in this case.

\par
In our theoretical analysis, we have not explicitly considered the effect of diffraction at the lenses. The theoretical predictions still match with the experimental observations. The reason for this can be understood as follows: the transverse dimensions of the nonlinear crystals (1 mm $\times$ 2 mm) constrain the maximum pump spot size and thus the maximally achievable momentum correlation (cf. Eq. \eqref{cond-prob-form}), i.e. how precisely $\qq_u$ is determined if the partner photon as an exact value of $\qq_d$. In our setup, 1 inch optics are used in the lens systems that map a transverse wave-vector $\qq_u$ $(\qq_d)$ to a point on the object (camera). Due to the diffraction limit imposed by the lenses \cite{goodman2005introduction}, the position of a point on the object (or camera) corresponding to a particular transverse wave vector has an uncertainty of  $\approx5$ $\mu$m. However, this uncertainty is much smaller than the uncertainty due to the imperfect correlation between $\qq_d$ and $\qq_u$; the latter corresponds to approximately $100$ $\mu$m for a pump waist of $200$ $\mu$m (roughly equivalent to the effect of a pump-spot sized aperture in the Fourier plane of the object). Thus, the diffraction limit imposed by the lenses is negligible in our experiment when compared to the resolution limit given by the imperfect momentum correlation.

\section{How to increase the resolution of future experiments} \label{sec:improvement}
In our analysis, we have considered experiments that use the scheme of Ref. \cite{lemos_quantum_2014}, in which nonlinear crystals are located
in the far field of the object and the momentum correlation between detected and undetected photons limits the resolution. In this case, the best resolution can be achieved if $\sigma/M$ in Eq. (\ref{PSFscaled}) is minimized. Equation (\ref{PSFscaled}) implies that in order to increase the resolution in this type of setup (at the cost of a decreased field-of-view (FOV)), it is beneficial to use lenses of short focal length in the undetected beam. Short focal lengths allow to decrease the distance between two points on an object, that correspond to two ``adjacent'' wave-vectors, i.e., to a fixed distance in the momentum space. A second alternative is decreasing the FOV by an external optical system.

The resolution can also be enhanced by decreasing the wavelength of the light illuminating the object. In order to increase the momentum correlation between the two photons, a large pump beam and hence nonlinear crystals with large transverse dimensions are
beneficial.
\par
A different approach could be to change the geometry of the setup in a way that the momentum correlation does not limit the resolution. This could be achieved, for example, by imaging the nonlinear crystals onto the object and the camera. In this case, the imaging system would rely on
photon pair correlations in the transverse position rather than in momentum. Therefore, the small size of the nonlinear crystals is expected to limit the FOV instead of the resolution.
\par
\section{Summary and Conclusions}\label{Sec:conclusions}
We have studied the resolution in Quantum Imaging with Undetected photons both theoretically and experimentally. Our experiment uses the scheme developed in Ref. \cite{lemos_quantum_2014}, in which object and camera are located in the far-field of the photon pair sources. 
In this case, we have identified the momentum correlation between detected and undetected photons as the dominant factor that limits the resolution.
The physical reason for this limit is that the spatial dimensions of the nonlinear crystals are much smaller than that of the lenses, which is a relevant case for practical applications.
\par
Our results show that in this regime, the width of the edge-spread function at the camera is determined by the wavelength of the detected light, whereas the resolution is dependent upon the wavelength of the undetected light illuminating the object. This shows that QIUP can be particularly promising in cases where the wavelength to which available cameras are sensitive is larger than the wavelength required for object illumination.
In a recent experiment, photon pairs \textemdash ~with one photon having wavelength in the range of x-ray and the other at optical wavelengths \textemdash ~have been generated \cite{schori2017parametric}. Therefore, it is certainly possible to design an experiment, where an object is illuminated by x-ray and the detected photon has optical wavelength. In such an experiment, it would in principle be possible to obtain x-ray resolution, using readily available cameras to detect only at optical wavelengths.
\par
Despite the fact that the results turned out to be analogous to that of quantum ghost imaging (QGI) enabled by momentum correlation, the theory and experiment of these two imaging schemes are fundamentally different. In QGI, the photon probing the object is detected and the image is acquired through coincidence measurement (post selection), whereas in QIUP the photon probing the object is never detected and the image is acquired through intensity (single-photon counting rate) measurement. Furthermore, QIUP is an interferometric scheme and QGI is not. Therefore, our finding that the resolution of QIUP based on momentum correlation is limited in the same way as QGI is quite remarkable, because it suggests that QIUP accesses the information encoded in two-photon correlations as efficiently as QGI.
\par
We expect that our results will enable future works leading to further improvement of this unique method of imaging and will thereby lead to a deeper understanding of its resolution limit in more general scenarios. It will be interesting to explore to what extent established super-resolution techniques can be adapted and to what extent the resulting implementations using other geometries or image acquisition methods will further increase the achievable resolution\footnote{While this manuscript was in submission, new articles were published  with interesting discussion on QIUP resolution~\cite{paterova2020hyperspectral,kviatkovsky2020microscopy,paterova2020quantum,paterova2021non,gilaberte2021video}.}.

\section*{Acknowledgements}
The authors thank Markus Gr\"afe and Marta Gilaberte for encouraging us to finish this project. This work was supported by the
Austrian Academy of Sciences (ÖAW) and the FWF project W 1210-N25 (CoQuS). JF and EAO also acknowledge ANID for the financial support (Becas de doctorado en el extranjero “Becas Chile”/ 2015 – No. 72160487 and 2016 – No. 72170402). RL acknowledges the funding by the National Science Centre (Poland) Project No. 2015/17/D/ST2/03471 and the Foundation for Polish Science within the FIRST TEAM project `Spatiotemporal photon correlation measurements for quantum metrology and super-resolution microscopy' co-financed by the European Union under the European Regional Development Fund (POIR.04.04.00-00-3004/17-00).
\printbibliography

@article{Klyshko-1997,
  title = {Interference effects in spontaneous two-photon parametric scattering from two macroscopic regions},
  author = {Burlakov, A. V. and Chekhova, M. V. and Klyshko, D. N. and Kulik, S. P. and Penin, A. N. and Shih, Y. H. and Strekalov, D. V.},
  journal = {Phys. Rev. A},
  volume = {56},
  issue = {4},
  pages = {3214--3225},
  numpages = {0},
  year = {1997},
  publisher = {American Physical Society},
  url ={https://doi.org/10.1103/PhysRevA.56.3214}
}

@article{bruschini2019single,
  title={Single-photon avalanche diode imagers in biophotonics: review and outlook},
  author={Bruschini, C. and Homulle, H. and Antolovic, I. M. and Burri, S. and Charbon, E.},
  %journal={Light: Science \& Applications},
  journal={Light Sci. Appl.},
  volume={8},
  number={1},
  pages={87},
  year={2019},
  url={https://doi.org/10.1038/s41377-019-0191-5},
  publisher={Nature Publishing Group}
}

@article{lubin2019quantum,
  title={Quantum correlation measurement with single photon avalanche diode arrays},
  author={Lubin, G. and Tenne, R. and Antolovic, I. M. and Charbon, E. and Bruschini, C. and Oron, D.},
  journal={Opt. Express},
  volume={27},
  number={23},
  pages={32863--32882},
  year={2019},
  url={https://doi.org/10.1364/OE.27.032863},
  publisher={Optical Society of America}
}

@article{patent,
	author= {Lemos, G. B. and Borish, V. and Ramelow, S. and Lapkiewicz, R. and Cole, G. D. and Zeilinger, A.},
	title={Quantum Imaging with Undetected Photons},
	journal={US Patent 9,557.262 B2; EP Patent 2 887 137 B1}}

@article{nomerotski2019imaging,
  title={Imaging and time stamping of photons with nanosecond resolution in Timepix based optical cameras},
  author={Nomerotski, A.},
  journal={Nucl. Instrum. Methods Phys. Res., Sect. A},
  volume={937},
  pages={26--30},
  year={2019},
  url={https://doi.org/10.1016/j.nima.2019.05.034},
  publisher={Elsevier}
}

@article{classen2017superresolution,
  title={Superresolution via structured illumination quantum correlation microscopy},
  author={Classen, A. and von Zanthier, J. and Scully, M. O. and Agarwal, G. S.},
  journal={Optica},
  volume={4},
  number={6},
  pages={580--587},
  year={2017},
  url={https://doi.org/10.1364/OPTICA.4.000580},
  publisher={Optical Society of America}
}

@article{kalashnikov2016infrared,
  title={Infrared spectroscopy with visible light},
  author={Kalashnikov, D. A. and Paterova, A. V. and Kulik, S. P. and Krivitsky, L. A.},
  journal={Nat. Photon.},
  volume={10},
  number={2},
  pages={98--101},
  year={2016},
  url={https://doi.org/10.1038/nphoton.2015.252},
  publisher={Nature Publishing Group}
}

@article{kutas2020terahertz,
  title={Terahertz quantum sensing},
  author={Kutas, M. and Haase, B. and Bickert, P. and Riexinger, F. and Molter, D. and von Freymann, G.},
  journal={Sci. Adv.},
  volume={6},
  number={11},
  pages={eaaz8065},
  year={2020},
  url={https://doi.org/10.1126/sciadv.aaz8065},
  publisher={American Association for the Advancement of Science}
}

@article{walborn2010spatial,
  title={Spatial correlations in parametric down-conversion},
  author={Walborn, S. P. and Monken, C. H. and P{\'a}dua, S. and Ribeiro, P. H. S.},
  journal={Phys. Rep.},
  volume={495},
  number={4-5},
  pages={87--139},
  year={2010},
  url={https://doi.org/10.1016/j.physrep.2010.06.003},
  publisher={Elsevier}
}

@article{Lantz-2005,
  title = {Spatially Noiseless Optical Amplification of Images},
  author = {Mosset, A. and Devaux, F. and Lantz, E.},
  journal = {Phys. Rev. Lett.},
  volume = {94},
  issue = {22},
  pages = {223603},
  numpages = {4},
  year = {2005},
  publisher = {American Physical Society},
  url = {https://doi.org/10.1103/PhysRevLett.94.223603}
  }

@article{moreau2019imaging,
  title={Imaging with quantum states of light},
  author={Moreau, P. A. and Toninelli, E. and Gregory, T. and Padgett, M. J.},
  journal={Nat. Rev. Phys.},
  volume={1},
  number={6},
  pages={367--380},
  year={2019},
  url={https://doi.org/10.1038/s42254-019-0056-0},
  publisher={Nature Publishing Group}
}

@article{Tsang-2016,
  title = {Far-Field Superresolution of Thermal Electromagnetic Sources at the Quantum Limit},
  author = {Nair, R. and Tsang, M.},
  journal = {Phys. Rev. Lett.},
  volume = {117},
  issue = {19},
  pages = {190801},
  numpages = {5},
  year = {2016},
  publisher = {American Physical Society},
  url = {https://doi.org/10.1103/PhysRevLett.117.190801}
  }

@article{Demkowicz-2018,
  title = {Beating the Rayleigh Limit Using Two-Photon Interference},
  author = {Parniak, M. and Bor\'owka, S. and Boroszko, K. and Wasilewski, W. and Banaszek, K. and Demkowicz-Dobrza\ifmmode \acute{n}\else \'{n}\fi{}ski, R.{}},
  journal = {Phys. Rev. Lett.},
  volume = {121},
  issue = {25},
  pages = {250503},
  numpages = {6},
  year = {2018},
  publisher = {American Physical Society},
  url = {https://doi.org/10.1103/PhysRevLett.121.250503}
}

@article{magana2019quantum,
  title={Quantum imaging and information},
  author={Maga{\~n}a-Loaiza, O. S. and Boyd, R. W.},
  journal={Rep. Prog. Phys.},
  volume={82},
  number={12},
  pages={124401},
  year={2019},
  url={https://doi.org/10.1088/1361-6633/ab5005},
  publisher={IOP Publishing}
}

@article{brida2010experimental,
  title={Experimental realization of sub-shot-noise quantum imaging},
  author={Brida, G. and Genovese, M. and Berchera, I. R.},
  journal={Nat. Photon.},
  volume={4},
  number={4},
  pages={227--230},
  year={2010},
  url={https://doi.org/10.1038/nphoton.2010.29},
  publisher={Nature Publishing Group}
}

@article{sabines2019twin,
  title={Twin-beam sub-shot-noise raster-scanning microscope},
  author={Sabines-Chesterking, J. and McMillan, A. R. and Moreau, P. A. and Joshi, S. K. and Knauer, S. and Johnston, E. and Rarity, J. G. and Matthews, J. C. F.},
  journal={Opt. Express},
  volume={27},
  number={21},
  pages={30810--30818},
  year={2019},
  url={https://doi.org/10.1364/OE.27.030810},
  publisher={Optical Society of America}
}

@article{garces2020quantum,
  title={Quantum-enhanced stimulated emission detection for label-free microscopy},
  author={Garces, G. T. and Chrzanowski, H. M. and Daryanoosh, S. and Thiel, V. and Marchant, A. L. and Patel, R. B. and Humphreys, P. C. and Datta, A. and Walmsley, I. A.},
  journal={Appl. Phys. Lett.},
  volume={117},
  number={2},
  pages={024002},
  year={2020},
  url={https://doi.org/10.1063/5.0009681},
  publisher={AIP Publishing LLC}
}

@article{schwartz2013superresolution,
  title={Superresolution microscopy with quantum emitters},
  author={Schwartz, O. and Levitt, J. M. and Tenne, R. and Itzhakov, S. and Deutsch, Z. and Oron, D.},
  journal={Nano Lett.},
  volume={13},
  number={12},
  pages={5832--5836},
  year={2013},
  url={https://doi.org/10.1021/nl402552m},
  publisher={ACS Publications}
}

@article{unternahrer2018super,
  title={Super-resolution quantum imaging at the Heisenberg limit},
  author={Untern{\"a}hrer, M. and Bessire, B. and Gasparini, L. and Perenzoni, M. and Stefanov, A.},
  journal={Optica},
  volume={5},
  number={9},
  pages={1150--1154},
  year={2018},
  url={https://doi.org/10.1364/OPTICA.5.001150},
  publisher={Optical Society of America}
}

@article{tenne2019super,
  title={Super-resolution enhancement by quantum image scanning microscopy},
  author={Tenne, R. and Rossman, U. and Rephael, B. and Israel, Y. and Krupinski-Ptaszek, A. and Lapkiewicz, R. and Silberberg, Y. and Oron, D.},
  journal={Nat. Photon.},
  volume={13},
  number={2},
  pages={116--122},
  year={2019},
  url={https://doi.org/10.1038/s41566-018-0324-z},
  publisher={Nature Publishing Group}
}

@article{lahiri2019nonclassical,
  title = {Nonclassicality of induced coherence without induced emission},
  author = {Lahiri, M. and Hochrainer, A. and Lapkiewicz, R. and Lemos, G. B. and Zeilinger, A.},
  journal = {Phys. Rev. A},
  volume = {100},
  issue = {5},
  pages = {053839},
  numpages = {7},
  year = {2019},
  publisher = {American Physical Society},
  url = {https://doi.org/10.1103/PhysRevA.100.053839}
}

@article{schori2017parametric,
  title={Parametric down-conversion of x rays into the optical regime},
  author={Schori, A. and B{\"o}mer, C. and Borodin, D. and Collins, S. P. and Detlefs, B. and Sala, M. M. and Yudovich, S. and Shwartz, S.},
  journal={Phys. Rev. Lett.},
  volume={119},
  number={25},
  pages={253902},
  year={2017},
  url={https://doi.org/10.1103/PhysRevLett.119.253902},
  publisher={American Physical Society}
}

@article{harms1977new,
	title={A new formulation of total unsharpness in radiography},
	author={Harms, A. A. and Zeilinger, A.},
	journal={Phys. Med. Biol.},
	volume={22},
	number={1},
	pages={70},
	year={1977},
	url={https://doi.org/10.1088/0031-9155/22/1/009},
	publisher={IOP Publishing}
}

@article{wiseman2000induced,
  title={Induced coherence with and without induced emission},
  author={Wiseman, H. M. and M{\o}lmer, K.},
  journal={Phys. Lett. A},
  volume={270},
  number={5},
  pages={245--248},
  year={2000},
  url={https://doi.org/10.1016/S0375-9601(00)00314-5},
  publisher={Elsevier}
}

@article{lemos_quantum_2014,
	title = {Quantum imaging with undetected photons},
	volume = {512},
	number = {7515},
	journal = {Nature},
	author = {Lemos, G. B. and Borish, V. and Cole, G. D. and Ramelow, S. and Lapkiewicz, R. and Zeilinger, A.},
	year = {2014},
	pages = {409--412},
	url={https://doi.org/10.1038/nature13586}
}

@article{lahiri2015theory,
  title={Theory of quantum imaging with undetected photons},
  author={Lahiri, M. and Lapkiewicz, R. and Lemos, G. B. and Zeilinger, A.},
  journal={Phys. Rev. A},
  volume={92},
  number={1},
  pages={013832},
  year={2015},
  url={https://doi.org/10.1103/PhysRevA.92.013832},
  publisher={APS}
}

@article{strekalov1995observation,
  title={Observation of two-photon “ghost” interference and diffraction},
  author={Strekalov, D. V. and Sergienko, A. V. and Klyshko, D. N. and Shih, Y. H.},
  journal={Phys. Rev. Lett.},
  volume={74},
  number={18},
  pages={3600},
  year={1995},
  url={https://doi.org/10.1103/PhysRevLett.74.3600},
  publisher={APS}
}

@article{pittman1995optical,
  title={Optical imaging by means of two-photon quantum entanglement},
  author={Pittman, T. B. and Shih, Y. H. and Strekalov, D. V. and Sergienko, A. V.},
  journal={Phys. Rev. A},
  volume={52},
  number={5},
  pages={R3429},
  year={1995},
  url={https://doi.org/10.1103/PhysRevA.52.R3429},
  publisher={APS}
}

@article{hochrainer2017quantifying,
  title={Quantifying the momentum correlation between two light beams by detecting one},
  author={Hochrainer, A. and Lahiri, M. and Lapkiewicz, R. and Lemos, G. B. and Zeilinger, A.},
  journal={Proc. Natl. Acad. Sci. USA},
  volume={114},
  number={7},
  pages={1508--1511},
  year={2017},
  url={https://doi.org/10.1073/pnas.1620979114},
  publisher={National Acad Sciences}
}

@article{lahiri2017twin,
  title={Twin-photon correlations in single-photon interference},
  author={Lahiri, M. and Hochrainer, A. and Lapkiewicz, R. and Lemos, G. B. and Zeilinger, A.},
  journal={Phys. Rev. A},
  volume={96},
  number={1},
  pages={013822},
  year={2017},
  url={https://doi.org/10.1103/PhysRevA.96.013822},
  publisher={APS}
}

@article{zou1991induced,
  title={Induced coherence and indistinguishability in optical interference},
  author={Zou, X. Y. and Wang, L. J. and Mandel, L.},
  journal={Phys. Rev. Lett.},
  volume={67},
  number={3},
  pages={318},
  year={1991},
  url={https://doi.org/10.1103/PhysRevLett.67.318},
  publisher={APS}
}

@article{wang1991induced,
  title={Induced coherence without induced emission},
  author={Wang, L. J. and Zou, X. Y. and Mandel, L.},
  journal={Phys. Rev. A},
  volume={44},
  number={7},
  pages={4614},
  year={1991},
  url={https://doi.org/10.1103/PhysRevA.44.4614},
  publisher={APS}
}

@article{monken1998transfer,
  title={Transfer of angular spectrum and image formation in spontaneous parametric down-conversion},
  author={Monken, C. H. and Ribeiro, P. H. S. and P{\'a}dua, S.},
  journal={Phys. Rev. A},
  volume={57},
  number={4},
  pages={3123},
  year={1998},
  url={https://doi.org/10.1103/PhysRevA.57.3123},
  publisher={APS}
}

@article{moreau2018resolution,
  title={Resolution limits of quantum ghost imaging},
  author={Moreau, P. A. and Toninelli, E. and Morris, P. A. and Aspden, R. S. and Gregory, T. and Spalding, G. and Boyd, R. W. and Padgett, M. J.},
  journal={Opt. Express},
  volume={26},
  number={6},
  pages={7528--7536},
  year={2018},
  url={https://doi.org/10.1364/OE.26.007528},
  publisher={Optical Society of America}
}

@article{bennink2002two,
  title={“Two-photon” coincidence imaging with a classical source},
  author={Bennink, R. S. and Bentley, S. J. and Boyd, R. W.},
  journal={Phys. Rev. Lett.},
  volume={89},
  number={11},
  pages={113601},
  year={2002},
  url={https://doi.org/10.1103/PhysRevLett.89.113601},
  publisher={APS}
}

@article{bennink2004quantum,
  title = {Quantum and Classical Coincidence Imaging},
  author = {Bennink, R. S. and Bentley, S. J. and Boyd, R. W. and Howell, J. C.},
  journal = {Phys. Rev. Lett.},
  volume = {92},
  issue = {3},
  pages = {033601},
  numpages = {4},
  year = {2004},
  publisher = {American Physical Society},
  url = {https://doi.org/10.1103/PhysRevLett.92.033601}
}

@article{Boyd-twocol-GI-2009,
  title = {Two-color ghost imaging},
  author = {Chan, K. W. C. and O'Sullivan, M. N. and Boyd, R. W.},
  journal = {Phys. Rev. A},
  volume = {79},
  issue = {3},
  pages = {033808},
  numpages = {6},
  year = {2009},
  publisher = {American Physical Society},
  url = {https://doi.org/10.1103/PhysRevA.79.033808}
}

@article{gatti2008quantumimaging,
title={Quantum imaging},
author={Gatti, A. and Brambilla, E. and Lugiato, L.},
journal={Progress in Optics},
volume={51},
pages={251--348},
year={2008}
}

@article{aspden2013epr,
  title={{EPR}-based ghost imaging using a single-photon-sensitive camera},
  author={Aspden, R. S. and Tasca, D. S. and Boyd, R. W. and Padgett, M. J.},
  journal={New J. Phys.},
  volume={15},
  number={7},
  pages={073032},
  year={2013},
  url={https://doi.org/10.1088/1367-2630/15/7/073032}
}

@book{goodman2005introduction,
  title={Introduction to Fourier optics},
  author={Goodman, J. W.},
  year={2005},
  publisher={Roberts and Company Publishers}
}

@book{born2013principles,
  title={Principles of optics},
  author={Born, M. and Wolf, E.},
  year={1999},
  edition={7},
  publisher={Cambridge University Press}
}

@book{boreman2001modulation,
  title={Modulation transfer function in optical and electro-optical systems},
  author={Boreman, G. D.},
  volume={4},
  year={2001},
  publisher={SPIE press Bellingham, WA}
}

@article{valles2018optical,
  title={Optical sectioning in induced coherence tomography with frequency-entangled photons},
  author={Vall{\'e}s, A. and Jim{\'e}nez, G. and Salazar-Serrano, L. J. and Torres, J. P.},
  journal={Phys. Rev. A},
  volume={97},
  number={2},
  pages={023824},
  year={2018},
  url={https://doi.org/10.1103/PhysRevA.97.023824},
  publisher={APS}
}

@article{paterova2018tunable,
  title={Tunable optical coherence tomography in the infrared range using visible photons},
  author={Paterova, A. V. and Yang, H. and An, C. and Kalashnikov, D. A. and Krivitsky, L. A.},
  journal={Quantum Sci. Technol.},
  volume={3},
  number={2},
  pages={025008},
  year={2018},
  url={https://doi.org/10.1088/2058-9565/aab567},
  publisher={IOP Publishing}
}

@article{paterova2020hyperspectral,
  title={Hyperspectral infrared microscopy with visible light},
  author={Paterova, A. V. and Maniam, S. M. and Yang, H. and Grenci, G. and Krivitsky, L. A.},
  journal={Sci. Adv.},
  volume={6},
  number={44},
  pages={eabd0460},
  year={2020},
  url={https://doi.org/10.1126/sciadv.abd0460},
  publisher={American Association for the Advancement of Science}
}

@article{kviatkovsky2020microscopy,
  title={Microscopy with undetected photons in the mid-infrared},
  author={Kviatkovsky, I. and Chrzanowski, H. M. and Avery, E. G. and Bartolomaeus, H. and Ramelow, S.},
  journal={Sci. Adv.},
  volume={6},
  number={42},
  pages={eabd0264},
  year={2020},
  url={https://doi.org/10.1126/sciadv.abd0264},
  publisher={American Association for the Advancement of Science}
}

@article{paterova2020quantum,
  title={Quantum imaging for the semiconductor industry},
  author={Paterova, A. V. and Yang, H. and Toa, Z. S. D. and Krivitsky, L. A.},
  journal={Appl. Phys. Lett.},
  volume={117},
  number={5},
  pages={054004},
  year={2020},
  url={https://doi.org/10.1063/5.0015614},
  publisher={AIP Publishing LLC}
}

@article{paterova2021non,
  title={Non-linear interferometry with infrared metasurfaces},
  author={Paterova, A. V. and Kalashnikov, D. A. and Khaidarov, E. and Yang, H. and Mass, T. W. W. and Paniagua-Dom{\'\i}nguez, R. and Kuznetsov, A. I. and Krivitsky, L. A.},
  journal={Nanophotonics},
  volume={10},
  number={6},
  pages={1775--1784},
  year={2021},
  url={https://doi.org/10.1515/nanoph-2021-0011},
  publisher={De Gruyter}
}

@article{gilaberte2021video,
  title={Video-Rate Imaging with Undetected Photons},
  author={Gilaberte Basset, M. and Hochrainer, A. and T{\"o}pfer, S. and Riexinger, F. and Bickert, P. and Le{\'o}n-Torres, J. R. and Steinlechner, F. and Gr{\"a}fe, M.},
  journal={Laser Photonics Rev.},
  volume={15},
  number={6},
  pages={2000327},
  year={2021},
  url={https://doi.org/10.1002/lpor.202000327},
  publisher={Wiley Online Library}
}
\onecolumn
\newpage
\appendix
\section{Details of the Setups}
\label{app-setupdetails}
In the first setup were used the set of wavelengths $\lambda_d=810$ nm, $\lambda_u=1550$ nm, and in front of the camera a long-pass filter 635 nm and an interference filter 810$\pm1$ nm. In the second setup were used the set of wavelengths $\lambda_d=842$ nm, $\lambda_u=780$ nm, and in front of the camera a long-pass filter 635 nm and an interference filter 842$\pm1$ nm. The optical distances in both setups were identical. The lenses to focus at the crystals were $LMC= 400,500,700,750,$ and $1000$ (all in mm). The imaging systems were built with the lenses $LP1=LP2=LU1=LU2=LD1=LD2=75$ mm, and $LC= 150$ mm. In both setups the non-linear crystals were PPKTP type-0 of dimensions 2x2x1 mm (lxwxh).    

\begin{figure}[htbp]
	\centering
	\includegraphics[width=1\linewidth]{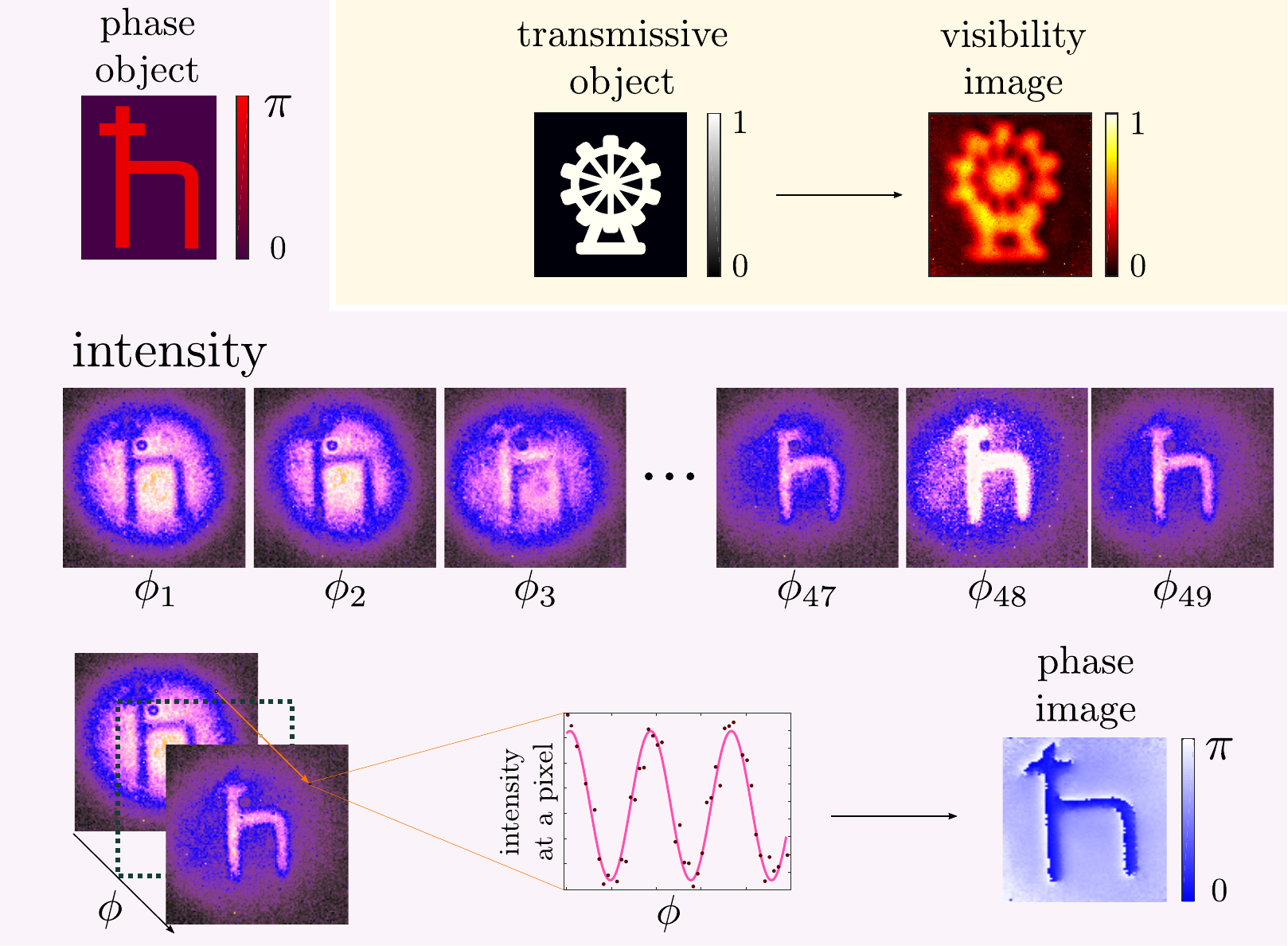}
	\caption{The imaging procedure for a phase object ($\hbar$). The image acquisition is the same for transmissive and phase objects (the upper row). The camera records the intensity of the detected beams while an interferometric phase $\phi$ is varied. The change in the interference patters with the varying phase is shown in the middle row. In total, 49 interference patterns were recorded, each one corresponding to a different value of $\phi = {\phi_1,...,\phi_{49}}$. One transverse momentum of the undetected beam photon acquires the information about one point of the object. That information is later recorded by a camera, where one transverse momentum of the detected photon is registered by one pixel on the camera. Thus, the information about each point of the object can be extracted independently by analyzing the recorded interference pattern pixel by pixel. In the lower row, a sinusoidal function is fitted over each interference pattern recorded by one pixel for different $\phi$. From this fit, the amplitude and phase corresponding to that pixel is extracted. The resulting amplitudes are used to construct a transmission image, while the relative phases are used to construct a phase image.}
	\label{fig:adquisition-appendix}
\end{figure}

The acquisition process is shown in Fig. \ref{fig:adquisition-appendix}. The $\hbar$ object produces a $\pi$ phase-shift at the wavelength $\lambda_u$=1550 nm. The acquisition time of each intensity image was 10 seconds. 

\section{Resolution for Phase Imaging}
\label{app-phaseimaging}
In this appendix, we justify that our main results are also true for a phase object. We have shown that the intensity at a point $x_c$ on the camera is given by Eq. (\ref{image-pattern}). For a phase object, we have $|T(\r_{\qq_u})|=1$, i.e. $T(\r_{\qq_u})=\exp[i\theta(\r_{\qq_u})]$. The resolution of a phase object can be studied by the use of the interference term, which now reduces to the form
\begin{align} \label{phase-image}
\mathcal{J}(\r_{\qq_d})=\sum_{\qq_u} p(\qq_u |
\qq_d)\cos(\theta(\r_{\qq_u})).
\end{align}
We find that the edge-spread function is characterized by the parameter 
\begin{align}\label{PSF-width-ph}
\sigma=\frac{f_c\lambda_d}{\sqrt{2}\pi w_p}.
\end{align}
Comparing this equation with Eq. (\ref{PSF-width}), we find that the quantity $\sigma$ (and consequently $\sigma/M$) has the same value for an absorptive and a phase edge. 
\par
\begin{figure*}[htbp]
	\centering
	\includegraphics[width=01\linewidth]
	{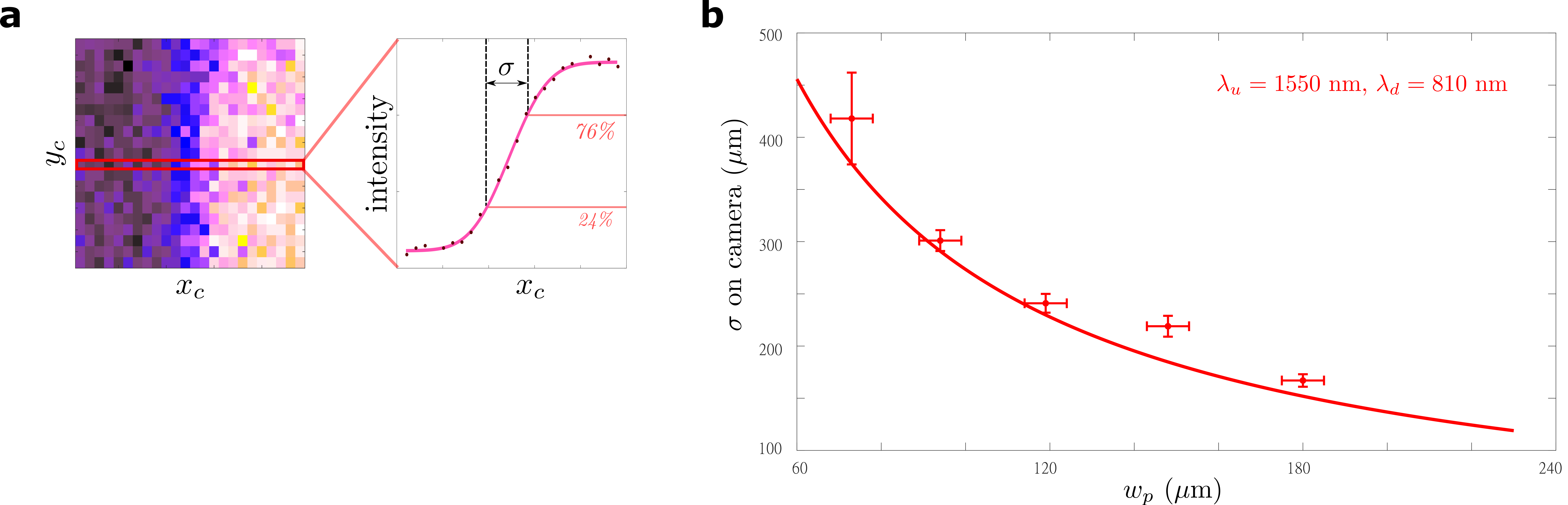}
    \caption{Corroboration of $\sigma$ for a phase object. In (a) is shown the method to obtain the characteristic distance $\sigma$ of the transition (24\% to 76\%) between maximum and minimum intensity. In (b) are presented the results of $\sigma$ for different pump waists. The results are in agreement with previous results showed for transmissive objects, where stronger momentum correlation (bigger $w_p$) leads to an improvement in resolution (smaller $\sigma$).  
    }
	\label{fig:phase-esf-resultati}
\end{figure*}

The experiment is done using the setup for which $\lambda_u=1550$ nm and $\lambda_d=810$ nm. To create a phase edge, we use a phase plate, TEM$_{01}$, which produces a $\pi$ phase difference for $1550$ nm between two parts of a beam cross-section. In a cross section of an intensity image showing constructive (or destructive) interference, is evaluated an error function to obtain the parameter $\sigma$ in different scenarios of momentum correlation. We evaluate the parameter $\sigma$ that characterizes the edge-spread function. In Fig. \ref{fig:phase-esf-resultati}, we illustrate the method and the experimental results. We conclude that the spatial resolution of both transmission and phase objects are limited by the imperfect momentum correlation of the photon pairs. 

\section{Details on the Test Target Evaluation}
\label{app-testtarge}
The minimum resolvable distance between two slit is a method for obtaining the resolvable power of an optical system. We used the object `1951 USAF resolution test chart', shown in Fig. 9(a). The object is formed by groups of six elements with three slits each (horizontal and vertical oriented). The slits have different width and separation through the object. In the method, we evaluated an image of two consecutive slits, with a waist of $w_p=201$ $\mu$m, and the wavelengths $\lambda_d=842$, and $\lambda_u=780$. In Fig. 9(c), we present the minimum resolvable distance between two slits adapted from Rayleigh's criterion \cite{born2013principles}. The threshold for two slit to be resolvable is $R=V_{max}/V_{min}=0.81$. An image of two consecutive slit that gives a coefficient $R \leq 0.81$, it is considered resolved. Additionally, we adapted the contrast transfer function. We define this function as $C=(V_{max}-V_{min})/(V_{max}+V_{min})$. In fig 9(b), we present the experimental results of $C$.  
 
\begin{figure*}[htbp]
	\centering
	\includegraphics[width=01\linewidth]
	{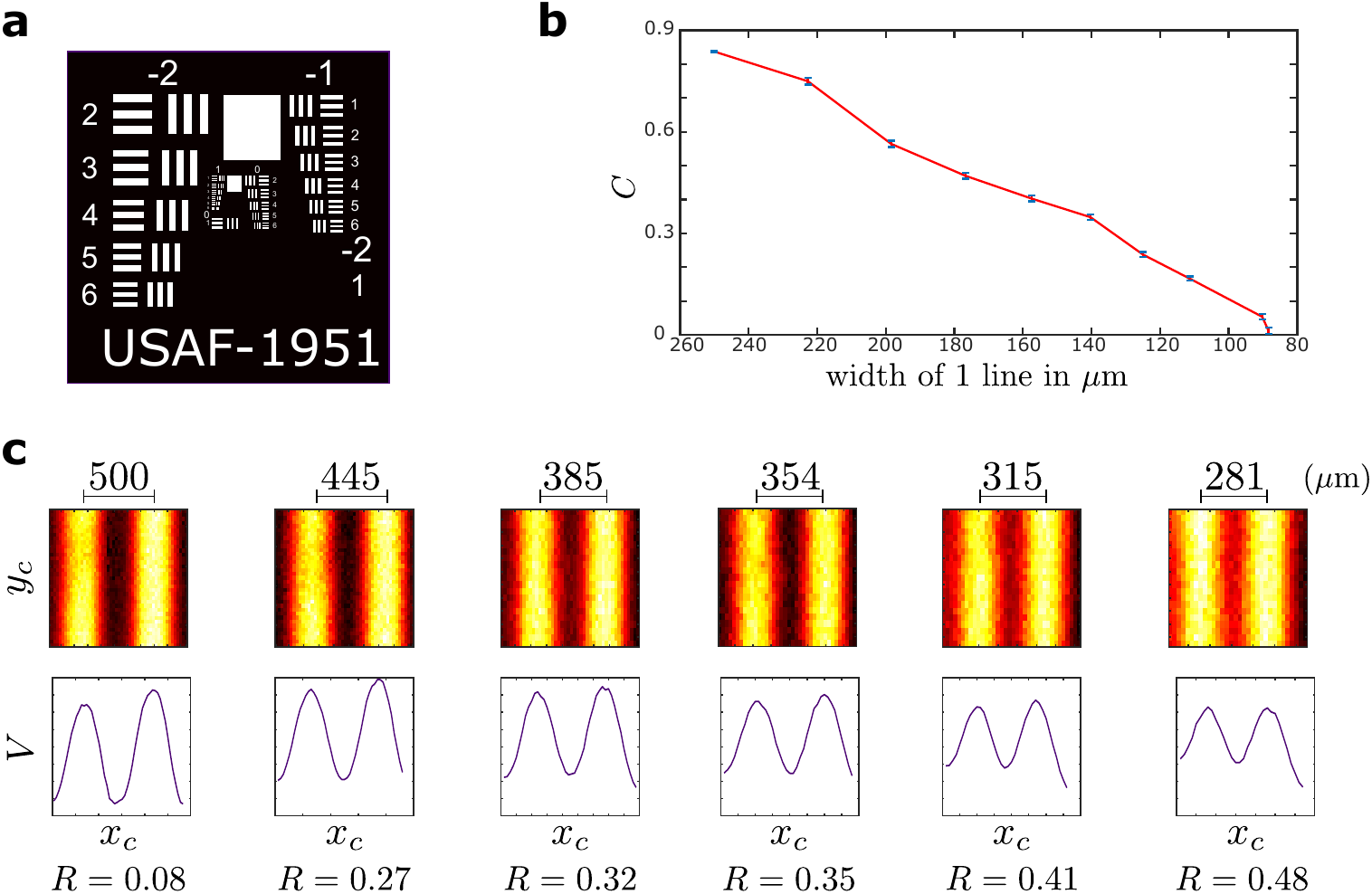}
    \caption{ Test target method and experimental results. The “USAF resolution test chart 1951” (a) had been used as the transmissive object in the experiment. The resolution of the same objects (USAF resolution test chart 1951, group 1, elements E1-E6) had been obtained by two different methods (b, c). In both methods,the maximum and minimum values are extracted from the visibility map. In (b), the contrast transfer function, given by $C=(V_{max}-V_{min})/(V_{max}+V_{min})$, is used as a resolution metric. A high value of $C$ corresponds to a higher resolution of the image. Thus, the object with a bigger slit separation has a higher value of $C$. With the slit separation decreasing, the values of $C$ become lower. In (c), the minimum resolvable distance of two slits, given by $R=V_{max}/V_{min}$, is used as another resolution metric. The value of $R$ is obtained over the cross-section of a two-slit image. Contrary to the previous method, a small value of $R$ corresponds to a higher resolution of the image. We adopted the Rayleigh's criterion to determine if an image is resolvable. A value $R \leq 0.81$ means that the image is resolvable. $y_c$ and $x_c$ correspond to the coordinates on the camera plane. The experimental beam parameters were: $w_p=201$ $\mu$m, $\lambda_d=842$ nm, and $\lambda_u=780$ nm.}
	\label{fig:test-target-procedure}
\end{figure*}

\end{document}